\documentclass[sigplan,screen]{acmart}

\AtBeginDocument{%
  }

\usepackage{url}
\usepackage{cleveref}
\usepackage{balance}
\usepackage{listings}
\usepackage{multirow}
\usepackage{microtype}
\usepackage[utf8]{inputenc}
\usepackage[caption=false]{subfig}

\usepackage{xspace}
\newcommand{\tool}{ADEPT\xspace}

\newcommand{\ie}{i.e.\xspace}
\newcommand{\eg}{e.g.\xspace}
\newcommand{\etc}{etc\xspace}

\newcommand{\ala}{à la\xspace}

\newcommand{\etal}{et al.\xspace}

\setcopyright{cc}
\setcctype{by}
\acmDOI{10.1145/3837729.3840488}
\acmYear{2026}
\copyrightyear{2026}
\acmISBN{979-8-4007-2910-2/2026/10}
\acmConference[SPLASH Companion '26]{Companion Proceedings of the 2026 ACM SIGPLAN International Conference on Systems, Programming, Languages, and Applications: Software for Humanity}{October 3--9, 2026}{Oakland, CA, USA}
\acmBooktitle{Companion Proceedings of the 2026 ACM SIGPLAN International Conference on Systems, Programming, Languages, and Applications: Software for Humanity (SPLASH Companion '26), October 3--9, 2026, Oakland, CA, USA}
\acmSubmissionID{splashcomp26demo-p14-p}
\received{2026-06-25}
\received[accepted]{2026-07-25}

\begin{document}

\title{ADEPT: A Unified Framework for Deep Learning Test Adequacy}

\author{Yidi Kao}
\correspondingauthor
\orcid{0009-0000-6264-2277}
\affiliation{%
  \institution{Auburn University}
  \department{Department of Computer Science and Software Engineering}
  \city{Auburn}
  \country{USA}
}
\email{yidik@auburn.edu}

\author{Shawn Burnham}
\orcid{0009-0002-1004-2725}
\affiliation{%
  \institution{Auburn University}
  \department{Department of Computer Science and Software Engineering}
  \city{Auburn}
  \country{USA}
}
\email{shawnburnham@auburn.edu}

\author{Tommi Rose Fahy}
\orcid{0009-0004-3239-6641}
\affiliation{%
  \institution{Auburn University}
  \department{Department of Computer Science and Software Engineering}
  \city{Auburn}
  \country{USA}
}
\email{tommirose.fahy@auburn.edu}

\author{Ali Ghanbari}
\orcid{0000-0003-1471-2546}
\affiliation{%
  \institution{Auburn University}
  \department{Department of Computer Science and Software Engineering}
  \city{Auburn}
  \country{USA}
}
\email{ghanbari@auburn.edu}

\begin{abstract}
Over the past decade, many test adequacy metrics have been proposed for deep learning that characterize test dataset adequacy from different perspectives, \eg, neuron activation behavior, latent feature coverage, decision-boundary exploration, \etc. 
However, these metrics are typically released as independent research prototypes with substantially different installation and preprocessing requirements, execution workflows, and configuration mechanisms. These complications make them quite difficult to reproduce, compare, and adopt in research work and practical deployment alike.

In this paper, we present the engineering details of \tool, a framework that integrates representative adequacy techniques, including neuron-coverage-based metrics, surprise adequacy, input distribution coverage, boundary coverage, and source- and model-level mutation score, under a consistent execution workflow. 
\tool provides a template-based metric interface with well-defined extension points for integrating new adequacy metrics.
Furthermore, it provides YAML-based configuration management, preprocessing-cache reuse, and structured result reporting, making it easy to use in any research and development workflows.
\tool is designed for researchers and practitioners who wish to reproduce and apply adequacy metrics without spending days or weeks implementing missing tooling or configuring disparate research prototypes.
A demo video is available at \url{https://aub.ie/ADEPT_video}.
\end{abstract}

%an \textbf{Ad}equacy \textbf{E}valuation framework for dee\textbf{P} learning \textbf{T}esting.
%\tool
% Implemented in Python and supporting Keras-based DNN models, \tool provides a unified and reproducible framework for executing and comparing heterogeneous adequacy metrics. It enables researchers and practitioners to readily assess test dataset adequacy throughout the DNN development lifecycle.
% These metrics have been used to support test dataset quality assessment, test prioritization, and have been widely studied on their own.
% Existing metrics 
%, activation trace distributions




%%
%% This command processes the author and affiliation and title
%% information and builds the first part of the formatted document.
\maketitle

\section{Introduction}\label{sec:intro}
Deep learning~\cite{bib:lecun2015deep} (DL) models are increasingly deployed in safety- and mission-critical domains such as autonomous driving, health care, and energy, warranting research to ensure their quality in terms of accuracy, robustness, or efficiency~\cite{bib:sarker2021deep}.
Among existing quality assurance techniques for DL models, testing remains one of the most widely used approaches in practice~\cite{bib:zhang2022machine,bib:hu2024test}.

However, good performance on a test dataset does not guarantee model accuracy, robustness, or efficiency on novel inputs. Therefore, there is a need to systematically assess the quality of test datasets themselves to understand how thoroughly they evaluate the model.
Additionally, by quantifying test dataset quality, one can compare datasets to each other and pick one dataset over another in a given scenario.
For example, practitioners often face large pools of candidate test inputs but limited labeling budgets and computational resources. As a result, they need methods to identify a small subset of inputs that can adequately evaluate the model, allowing them to focus their effort on that subset.
Consequently, over the past decade, researchers have proposed numerous DNN test adequacy metrics to assess test-set quality and support applications such as test dataset prioritization and evaluation.
These metrics include, but are not limited to, neuron coverage~\cite{bib:pei2017deepxplore} and its variations~\cite{bib:ma2018deepgauge}, surprise adequacy~\cite{bib:kim2018guiding}, input distribution coverage~\cite{bib:dola2023input}, decision-boundary coverage~\cite{bib:liu2022deepboundary}, and source- and model-based mutation score~\cite{bib:hu2019deepmutation++,bib:humbatova2021deepcrime}.

Despite the large number of adequacy metrics, existing implementations remain highly fragmented. When the implementation is available, different approaches rely on substantially different preprocessing pipelines, execution workflows, and setup requirements.
For example, some metrics based on neuron coverage, like KMNC~\cite{bib:ma2018deepgauge}, require activation profiling over training data, surprise adequacy relies on activation trace extraction and density estimation, input distribution coverage depends on separately trained encoder models, and mutation testing typically requires generating and executing a large collection of mutants.
As a result, reproducing, comparing, and deploying these adequacy metrics in practice often requires substantial effort, either to re-implement unavailable metrics or to configure and run available ones.
Case in point, recent interviews with industrial DNN practitioners reveal that existing testing metrics and tools are difficult to use in practice due to heterogeneous workflows, high adoption barriers, limited tool integration, and the absence of publicly available implementations for some techniques~\cite{bib:you2025navigating}.

To address this problem, we present \tool, an extensible framework for running DL test adequacy metrics.
\tool provides a unified command-line interface for executing representative adequacy metrics, including neuron coverage and its variants, surprise adequacy, input distribution coverage, decision boundary coverage, and source-/model-level mutation score.
The framework is implemented in Python and supports Keras-based DL models.
Its unified architecture implements each adequacy metric as an independent module with consistent interfaces for preprocessing and scoring while preserving metric-specific execution logic, allowing users to easily integrate new metrics into the framework.
To improve efficiency under repeated evaluations, the framework also supports reusable caching of intermediate artifacts generated during preprocessing, which reduces redundant computation when metrics are repeatedly executed.
% To support practical usage, the framework provides command-line interfaces to evaluate test subsets within a consistent workflow.
In this paper, we present the engineering details of \tool and show how it can be used.

\tool is publicly available at \url{https://zenodo.org/records/21682100} and a demo video can be found at \url{https://aub.ie/ADEPT_video}.
% We demonstrate its usefulness through representative adequacy metrics and evaluation workflows drawn from practical DL testing scenarios.

% Our contributions are summarized as follows:

% \begin{itemize}

% \item We present \tool, a one-click, unified framework for DNN test dataset adequacy assessment that simplifies the execution and comparison of representative adequacy metrics through a consistent evaluation workflow.

% \item We integrate representative adequacy techniques with substantially different preprocessing requirements into a common execution framework, enabling consistent evaluation and comparison across neuron-coverage-based, surprise-adequacy-based, latent-distribution-based, decision-boundary-based, and mutation-score-based approaches.

% \item We provide standardized interfaces for metric execution, result collection, and runtime reporting, reducing engineering effort and improving the reproducibility and usability of DNN test dataset adequacy assessment.

% \end{itemize}
\section{\tool Overview}\label{sec:overview}
To address implementation fragmentation, metric-specific execution workflows, and the lack of public code for some representative metrics, we design \tool, a unified framework for assessing the adequacy of DL test datasets.

\begin{figure*}[t]
\centering
\includegraphics[width=0.75\linewidth]{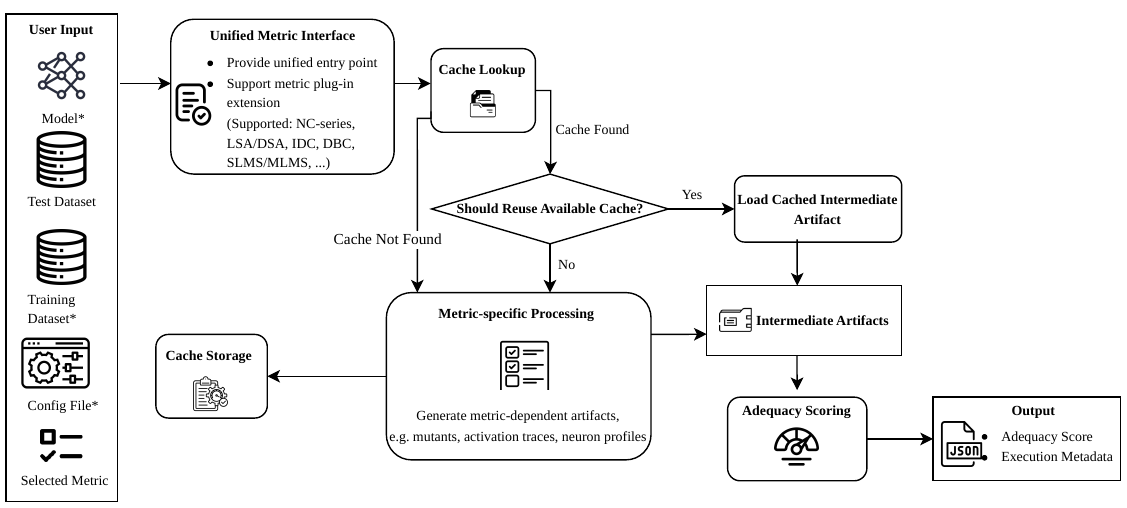}
\caption{Overview of \tool. Inputs marked with an asterisk (*) are optional and are required only by specific metrics.}
\label{fig:framework}
\end{figure*}

Figure~\ref{fig:framework} presents the overall architecture of \tool, which integrates representative adequacy metrics under a common workflow while preserving their metric-specific processing requirements.

The framework consists of four main components: (1) a template-based metric interface that provides access to representative adequacy metrics and enables new metrics to be integrated through well-defined extension points, (2) metric-specific processing modules that perform the preprocessing required by each adequacy technique and generate the corresponding intermediate artifacts, (3) a cache management component that stores and retrieves reusable intermediate artifacts across repeated evaluations, and (4) an adequacy scoring component that computes the final adequacy score for a given test dataset and generates a final report containing the score and execution metadata.
% To avoid redundant computation, \tool also maintains reusable cached artifacts, such as neuron profiles, activation traces, and generated mutants, which can be reused across repeated evaluations.
% \ali{This last sentence is something that you could perhaps explain and brag about in the Introduction section.}
% \ali{Please also clarify here whether Unified Metric Interface is extendable, \ie, if researchers can add new metrics to it with relative ease. This should be explained in Section~\ref{sec:overview:interface} as well.}

\subsection{Unified Metric Interface}\label{sec:overview:interface}% and Supported Metrics
The metric interface provides a unified entry point for executing different DL test dataset adequacy metrics.
Each metric is implemented as a plug-in module that follows the same framework-level abstraction while preserving its own processing and scoring logic.
This design allows \tool to support different adequacy techniques without requiring users to coordinate their different execution requirements manually.
New metrics can also be incorporated by implementing a metric module that follows the same template-based interface.
% \ali{And make the framework extendable?}

\tool integrates several representative DL test adequacy techniques that have been proposed in prior works. 
Currently, the framework supports neuron coverage and its variations (NC-series), two flavors of surprise adequacy (LSA and DSA), input distribution coverage (IDC), deep boundary coverage (DBC), and source- and model-level mutation score (SLMS and MLMS).
% \ali{Be consistent with these newly introduced abbreviations in Figure~\ref{fig:framework} and throughout the paper.}

NC-series metrics measure adequacy based on neuron activation behavior observed during model execution.
NC and Top-k Neuron Coverage (TKNC) assess whether individual neurons are activated above a threshold or ranked among the top-k most active neurons for a given input. 
% \ali{(have you explained what TKNC stand for? If so, you're good, if not it must be defined.)}
$k$-multisection Neuron Coverage (KMNC), Neuron Boundary Coverage (NBC), and Strong Neuron Activation Coverage (SNAC) characterize coverage relative to the activation ranges observed on the training set.

% \ali{Once introduced an abbreviation, please go with that abbreviation instead of repeating the full name every time.}
LSA and DSA assess test adequacy by measuring how surprising test inputs are relative to the training data based on their activation traces.

IDC measures adequacy in a learned latent feature space.
It employs a variational autoencoder ~\cite{bib:kingma2014autoencoding,bib:higgins2017betavae} (VAE) to construct a compact representation of the input distribution and applies combinatorial interaction testing to assess feature diversity within a test dataset.

DBC measures test adequacy from a decision-space perspective, focusing on how test inputs cover the model’s decision boundaries.

SLMS and MLMS evaluate test datasets based on their ability to distinguish mutant models from the original model.
The underlying intuition is that more effective test datasets should be capable of exposing behavioral differences introduced by injected faults.
\tool supports both MLMS, which applies mutation operators directly to trained DL models, and SLMS, which introduces faults before the training process to generate mutants.

Each adequacy metric is accessible through a consistent execution interface. Users specify the test dataset, target model, training dataset if needed, and adequacy metric, while \tool automatically invokes the corresponding processing pipeline. This reduces the effort required to use different adequacy techniques.

\subsection{Metric-Specific Processing}
Different adequacy metrics require different forms of preprocessing before adequacy scores can be computed.
To support metric-specific requirements and improve extensibility, \tool implements each metric's preprocessing workflow as a dedicated processing module.

For NC-based metrics that rely on training-set activation statistics, \ie, KMNC, NBC, and SNAC, \tool performs neuron profiling on the training dataset.
For LSA and DSA, activation traces are extracted and used for surprise estimation.
IDC relies on pre-trained VAE models to construct latent feature representations, while DBC performs decision-boundary extraction to characterize model behavior near classification boundaries.
\tool carries out MLMS \ala DeepMutation++~\cite{bib:hu2019deepmutation++} and SLMS \ala DeepCrime~\cite{bib:humbatova2021deepcrime}.
% For Mutation Score, \tool supports both model-level and source-level mutation workflows using configurable mutation operators. In the model-level workflow, \tool computes mutation scores during mutant generation by reusing execution results immediately, avoiding the need to perform mutation testing as a separate post-processing step after all mutants have been generated.

Although these preprocessing procedures differ substantially across adequacy metrics, \tool automatically invokes the corresponding processing module and generates the intermediate artifacts needed for score computation.
This design separates metric-specific implementation details from the external execution workflow and reduces the effort required to reproduce and compare different adequacy techniques.
% \vspace{-0.2cm}

\subsection{Cache Management}
% \ali{This is a major selling point that must be highlighted in the introduction, demo video, and if possible even in the abstract.}
Many adequacy metrics require intermediate artifacts that are costly to generate but can be reused across repeated evaluations.
To avoid redundant computation, \tool provides a cache management component that stores and retrieves these reusable artifacts generated during metric-specific processing.

Examples of cached artifacts include neuron profiles for some NC-series metrics, activation traces for LSA/DSA, decision-boundary models for DBC, and generated mutant models for SLMS and MLMS.
Cached artifacts are associated with the target model, dataset, metric, and creation time.
When compatible cached artifacts are available, \tool allows users to choose either to reuse the existing artifacts and proceed directly to scoring, or to regenerate them by invoking the corresponding metric-specific processing module.
This mechanism reduces the cost of repeated execution while preserving reproducibility and experimental traceability.

% \vspace{-0.2cm}
\subsection{Adequacy Scoring and Reporting}
The adequacy scoring component uses the artifacts produced by the processing stage to compute the final score according to the selected metric.
Coverage-based metrics report the proportion of covered adequacy elements, LSA/DSA derives scores from activation trace distributions, IDC measures latent feature interaction coverage, DBC quantifies decision-boundary coverage, and SLMS/MLMS reports the classic mutation score~\cite{bib:hu2019deepmutation++}, computed as the proportion of killed mutants over all generated mutants, i.e., $\text{MS} = \frac{|\text{killed mutants}|}{|\text{all mutants}|}$. 
% \ali{either statistically like DeepCrime, using DeepMutation formula, or as classic proportion of killed mutants.}

For each run, \tool produces the metric adequacy score together with lightweight execution metadata.
These metadata include preprocessing time, scoring time, and cache utilization information.
The resulting report allows users to inspect the adequacy score, understand the computational cost of each metric, and compare multiple adequacy techniques under a consistent evaluation setting.

Together, these components provide a unified workflow for running, inspecting, comparing, and extending diverse DL test adequacy metrics.
\section{\tool Usage}
% \ali{Make sure all links to YouTube and GitHub are shortened \via \url{aub.ie}.}
\tool is implemented in Python and can be used through a command-line interface.
After checking out the source code from the GitHub repository~\cite{adept}, users can create the required Python environment using the provided \texttt{environment.yml} file.
The framework currently accepts NumPy (\texttt{.npy}) datasets as input and supports Keras-based DNN models for metrics that require access to the tested model.

Once the environment is set up, users can execute an adequacy metric by specifying the target model, dataset name, test inputs, optional training data, and a metric-specific configuration file.
The following command illustrates the general usage pattern:
% \vspace{-0.1cm}
\begin{small}
\begin{verbatim}
python -m adequacy.cli run \
  --metric [metric-name] \
  --model path/to/model.keras \
  --dataset [dataset-name] \
  --test-data path/to/test_inputs.npy \
  --test-labels path/to/test_outputs.npy \
  --config configs/[metric-name].yaml
\end{verbatim}
\end{small}
% \vspace{-0.1cm}
For metrics that require additional inputs, such as training data for neuron profiling, activation-trace extraction, or mutation-score computation, \tool checks the required arguments before execution and reports missing inputs to the user.
\begin{table*}[t]
    \centering
    \caption{Representative configuration parameters used by \tool. 
    % \ali{Adopt SLMS and MLMS appropriately in this table AND in the \textit{source code}.}
    }
    \label{tab:config}
     \setlength{\tabcolsep}{4pt}
    \renewcommand{\arraystretch}{0.8}
    \scriptsize
    \begin{tabular}{llllp{5.5cm}}
        \toprule
        \textbf{Metric} & \textbf{Parameter} & \textbf{Metric group} & \textbf{Description} \\
        \midrule
        KMNC, NBC, SNAC & \texttt{train\_fraction} & \multirow{2}{*}{NC-series} & Fraction of training data used for neuron profiling \\
        All NC-series metrics & \texttt{exclude\_layer}  & & Layer name substrings excluded from profiling \\
        \midrule
        LSA, DSA   & \texttt{layer\_names}   & \multirow{2}{*}{SA} & Target layers for AT extraction; auto-inferred if omitted \\
        LSA                  & \texttt{var\_threshold} &  & Variance threshold for AT dimension filtering before KDE fitting \\
        \midrule
        DBC       & \texttt{beta}           & \multirow{2}{*}{DBC} & Convergence threshold for bisection-based boundary search \\
        DBC                  & \texttt{max\_per\_pair} &  & Maximum boundary points searched per class pair \\
        \midrule
        IDC       & \texttt{vae\_model\_dir} & \multirow{2}{*}{IDC} & Path to pre-trained VAE model directory (\textit{required}) \\
        IDC                  & \texttt{ways}            &  & $t$-way interaction strength for combinatorial coverage \\
        \midrule
        MLMS  & \texttt{num\_mutants} & \multirow{2}{*}{MS} & Number of mutants to generate \\
        MLMS                        & \texttt{mode}         &  & Execution mode: \texttt{generate\_and\_score} or \texttt{score\_from\_cache} \\
        \midrule
        SLMS & \texttt{mutation\_operators} & \multirow{2}{*}{MS} & List of mutation operators applied to training pipeline \\
        SLMS                        & \texttt{cache\_mutants}       &  & Whether to cache mutant models across runs \\
        \bottomrule
    \end{tabular}
\end{table*}

Metric-specific options are specified through YAML configuration files.
Table~\ref{tab:config} summarizes representative configuration parameters for different adequacy metrics in \tool.
These configuration files allow users to adjust processing and scoring behavior without modifying the source code.
If a configuration file is not provided, \tool falls back to default values defined in each metric module, enabling users to apply the supported adequacy metrics under common settings. These default configurations are documented in the GitHub repository~\cite{adept} README.

During processing, \tool stores reusable intermediate artifacts in the cache directory.
% These artifacts may include neuron profiles, activation traces, decision-boundary models, and generated mutant models.
When compatible cached artifacts are available, the framework prompts users to choose whether to reuse the existing artifacts or regenerate them; this choice determines whether the framework skips metric-specific processing and proceeds directly to scoring.
% This mechanism allows users to avoid repeated expensive preprocessing while still keeping the execution process explicit and reproducible.

After each run, \tool generates a structured JSON report containing the final adequacy score, important configuration, preprocessing time, scoring time, and cache utilization information. These outputs allow users to inspect each run, reproduce adequacy evaluations, and compare multiple metrics under a consistent workflow.

% \vspace{-0.5cm}
\section{Related Works}
% \ali{Perhaps you can provide a more thorough related work and cite things like MC/DC-style adequacy by Sun \etal. and Ma \etal's DeepCT and so on.}

DL test adequacy metrics have been widely studied and can be organized according to classical software testing coverage perspectives~\cite{bib:ammann2008introduction}, depending on the type of signal used to assess test effectiveness. 

\textbf{Structural adequacy metrics.}
Structural-based methods evaluate test adequacy using internal structural features and activation dynamics of the network. NC by Pei \etal~\cite{bib:pei2017deepxplore} and its variants in DeepGauge~\cite{bib:ma2018deepgauge} measure how test inputs activate neurons or cover activation patterns within the network. Beyond single-neuron activation, interaction-based criteria such as DeepCT~\cite{bib:ma2019deepct} and MC/DC-style adequacy by Sun \etal~\cite{bib:sun2018testing} extend structural testing principles to neuron interaction relationships, capturing complex inter-dependencies among layer activations.

\textbf{Input-space adequacy metrics.}
Input-space adequacy metrics estimate test adequacy by characterizing how test inputs cover or deviate within input-related representation spaces.
SA~\cite{bib:kim2018guiding} evaluates test inputs by measuring their deviation from training data behavior in activation space.
These methods quantify input novelty using activation traces and statistical distance measures, reflecting how ``surprising'' a test input is with respect to previously observed behaviors.
IDC~\cite{bib:dola2023input} assesses test adequacy in learned feature spaces, employing VAE~\cite{bib:kingma2014autoencoding,bib:higgins2017betavae} and combinatorial interaction testing over latent features to measure coverage of input feature interactions.

\textbf{Boundary-based metrics.}
Boundary-based approaches evaluate test adequacy using model uncertainty and decision-region characteristics. DeepBoundary by Liu \etal~\cite{bib:liu2022deepboundary} characterizes test dataset adequacy by modeling the distribution of inputs with respect to the distance to the decision boundary, capturing how close test inputs are to boundary regions.

\textbf{Mutation-based metrics.}
Mutation-based approaches evaluate test dataset effectiveness by introducing model-level or source-level mutations and measuring whether the test dataset can distinguish the original model from its mutated versions.
Representative methods include DeepMutation++~\cite{bib:hu2019deepmutation++}, DeepMutation~\cite{bib:ma2018deepmutation}, and DeepCrime~\cite{bib:humbatova2021deepcrime}.

Existing available implementations, such as DeepXplore~\cite{bib:pei2017deepxplore}, IDC~\cite{bib:dola2023input}, and DeepCrime~\cite{bib:humbatova2021deepcrime}, only focus on specific metrics rather than providing a unified workflow. More broadly, many proposed adequacy metrics remain difficult to run in practice: some lack publicly available implementations, while others depend on broken packages, limited maintenance, or complicated environment setup. A recent empirical study by You \etal~\cite{bib:you2025navigating} further highlights these practical challenges. Practitioners report difficulties in selecting and applying adequacy metrics due to a lack of unified tooling and inconsistent workflows across implementations.

In contrast, \tool provides a unified and extensible framework for DL test adequacy evaluation, enabling consistent execution and reuse of representative metrics under a common workflow.

% \vspace{-0.5cm}
\section{Tool Availability}
% \ali{Make sure all links are shortened \via \url{aub.ie}.}
% The source code, documentation, example configurations, and usage instructions for
\tool are publicly available on Zenodo:
\url{https://zenodo.org/records/21682100}.
A demo video is also available on YouTube: \url{https://aub.ie/ADEPT_video}

% A companion demonstration video showcasing the framework architecture, configuration process, and metric execution workflow is available at: .

% \vspace{-0.5cm}
\section{Conclusions and Future Work}
This paper presents \tool, a unified framework for DL test dataset adequacy assessment.
The framework integrates representative adequacy metrics in a consistent execution workflow while preserving their metric-specific processing requirements.
By providing extensible mechanisms for configuration, metric processing, and reporting, \tool simplifies the execution, comparison, and deployment of diverse adequacy techniques.
Additionally, by minimizing the engineering effort required to evaluate existing metrics, \tool enables researchers and practitioners to easily integrate test adequacy assessment into the development lifecycle of the DL system.

% As DL models continue to evolve through retraining, fine-tuning, and deployment optimization, practitioners are increasingly faced with large test datasets and limited labeling resources. In such settings, the assessment of the adequacy can provide valuable guidance for selecting, prioritizing, and evaluating test datasets. By reducing the engineering effort required to configure and execute existing adequacy metrics, \tool enables researchers and practitioners to more easily incorporate adequacy assessment into the DL development lifecycle.

% For future work, we plan to further improve the extensibility of the framework by introducing a lightweight plug-in mechanism that allows the newly proposed adequacy metrics to be integrated with minimal implementation effort.
As new adequacy techniques continue to emerge, we envision \tool as a continuously evolving platform that facilitates the reproducible evaluation and practical adoption of DL test adequacy assessment methods.
As future work, we plan to incorporate more metrics in the framework.

\section*{Acknowledgments}
This research is partially supported by the NSF grant \#2446393.
Any opinions, findings, and conclusions or recommendations expressed in this material are those of the authors and do not necessarily
reflect the views of the NSF.

\bibliographystyle{ACM-Reference-Format}
\bibliography{main}

\end{document}